\documentstyle[11pt,newpasp,twoside,epsf]{article}
\def\edcomment#1{\iffalse\marginpar{\raggedright\sl#1\/}\else\relax\fi}
\reversemarginpar

\begin{document}
\title{The PSR~J0514$-$4002A binary system in NGC 1851}
 \author{P. C. Freire}
\affil{Arecibo Observatory, HC 3 Box 53995, Arecibo, PR 00612, USA}
\author{Y. Gupta}
\affil{National Centre for Radio Astrophysics, P.O. Bag 3, Ganeshkhind, Pune 411007, India}
\author{S. M. Ransom}
\affil{McGill University, 3600 University St., Montreal, QC H3A 2T8
 Canada}
\author{C. H. Ishwara-Chandra}
\affil{National Centre for Radio Astrophysics, P.O. Bag 3, Ganeshkhind, Pune 411007, India}

\begin{abstract}
Using the Giant Metrewave Radio Telescope (GMRT), we have discovered
PSR~J0514$-$4002A, a binary millisecond pulsar in the globular cluster
NGC~1851. This pulsar has a rotational period of 4.99 ms and the most
eccentric pulsar orbit yet found: $e = 0.89$. The orbital period
is 18.8 days, and companion has a minimum mass of 0.9~M$_{\odot}$; its
nature is presently unclear. After accreting matter from a low-mass
stellar companion, this pulsar exchanged it for its more massive
present companion. This system presents the strongest evidence to date
of such a process.
\end{abstract}

\section{Introduction}

Since 1987, several globular cluster surveys (see review by F. Camilo
and papers by A. Possenti, S. Ransom and J. Hessels, these proceedings) have
confirmed that most of the binary millisecond pulsars (MSPs) in
globular clusters (GCs) have low-mass white dwarf companions and
nearly circular orbits, as observed in the Galactic disk. This is an
important confirmation of the evolutionary scenarios proposed by
Alpar~et~al.~(1982) for the formation of MSPs.
 
In GCs, {\em exchange encounters}, which only have a significant
probability of occurring in dense stellar environments, occasionally
exchange one of the components of a binary system with a typically
more massive star.  The exchanges may occur during encounters with
either other binaries or with isolated stars.  In GCs such encounters
can place isolated neutron stars into binaries with a main sequence
(MS) star which eventually evolves, ``recycles'' the neutron star, and
finally forms a MSP$-$WD binary system.  Such a process explains the
anomalously large numbers of MSPs in GCs (by mass) when compared to
the Galaxy. If the companion to the low-mass MS star is a previously
recycled neutron star, we observe ``irregular'' eclipsing binary
pulsars (see review by P. Freire).

In this paper we present some preliminary results on a new, unique
binary millisecond pulsar, PSR~J0514$-$4002A (henceforth NGC~1851A) in
the globular cluster NGC~1851. This has been found in a
new, sensitive 327-MHz survey of globular clusters carried out using
the Giant Metrewave Radio Telescope (GMRT), at Khodad near Pune, India.

\begin{figure}
\setlength{\unitlength}{1in}
\begin{picture}(0,2.8)
\put(0.1,-2.2){\includegraphics{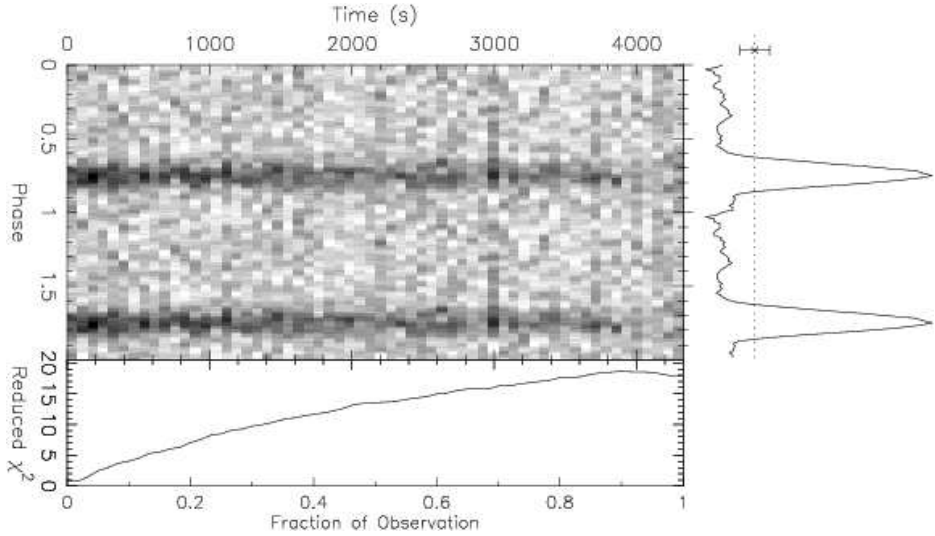}}
\end{picture}
\caption{Discovery observation for NGC~1851A, in the globular cluster
NGC~1851. Pulsed emission is persistent throughout the whole scan. The
pulse profile is rather narrow (right), corresponding to the time
resolution of the system for this DM.}
\end{figure}

\section{The GMRT 327-MHz survey and the discovery of NGC1851A}

Our use of a low radio frequency for the present pulsar survey is to
be contrasted with most recent pulsar surveys, which are carried at
higher radio frequencies. This survey is aimed at faint pulsars
with steep spectra, which are unlikely to be detectable in a
high-frequency survey. This particular survey benefits from the
large gain of the central array of the GMRT, 4.6~K/Jy when used in the
phased array mode of the Array Combiner (Gupta et al. 2000, Prabu
1997). This produces a beam on the sky with a diameter of about 3
arcminutes. The number of spectral channels across the available
16-MHz band is 256. The sampling time used is 258$\,\mu$s, each
observation consisting of a pair of 72-minute scans containing
2$^{24}$ samples. After this time, the 14-antenna central array is
re-phased using a reference source.
 
We observed a set of 16 GCs in February 2003. The data were written to
tape at the GMRT and taken to McGill University, where they were
processed using the BORG (a 104-node Beowulf cluster available there
for pulsar processing) running the PRESTO software package (Ransom 2001).
One of the GCs observed was NGC~1851. Its distance ($D$) from the Sun
is about 12.6~kpc ({Cassisi}, {De Santis}, \&
{Piersimoni} 2001), and its Galactic coordinates are
$l = 244.51^\circ, b = -35.04^\circ$ (Harris 1996)\footnote{See
http://physwww.physics.mcmaster.ca/$\sim$harris/mwgc.dat
for an updated version of the table of globular cluster parameters
presented in this paper}. It is
a relatively bright globular cluster ($M_v\,=\,-8.33$) with a very
condensed core ($c = log(r_t/r_c) = 2.32$, where $r_t$ and
$r_c$ are the tidal and core radii). It is among the ten clusters in
the Galaxy with the highest central luminosity density
($\rho_0\,\simeq\,2 \times 10^5 \rm L_{\odot} pc^{-3}$).

\begin{figure}
\setlength{\unitlength}{1in}
\begin{picture}(0,3.5)
\put(-0.3,-0.5){\includegraphics{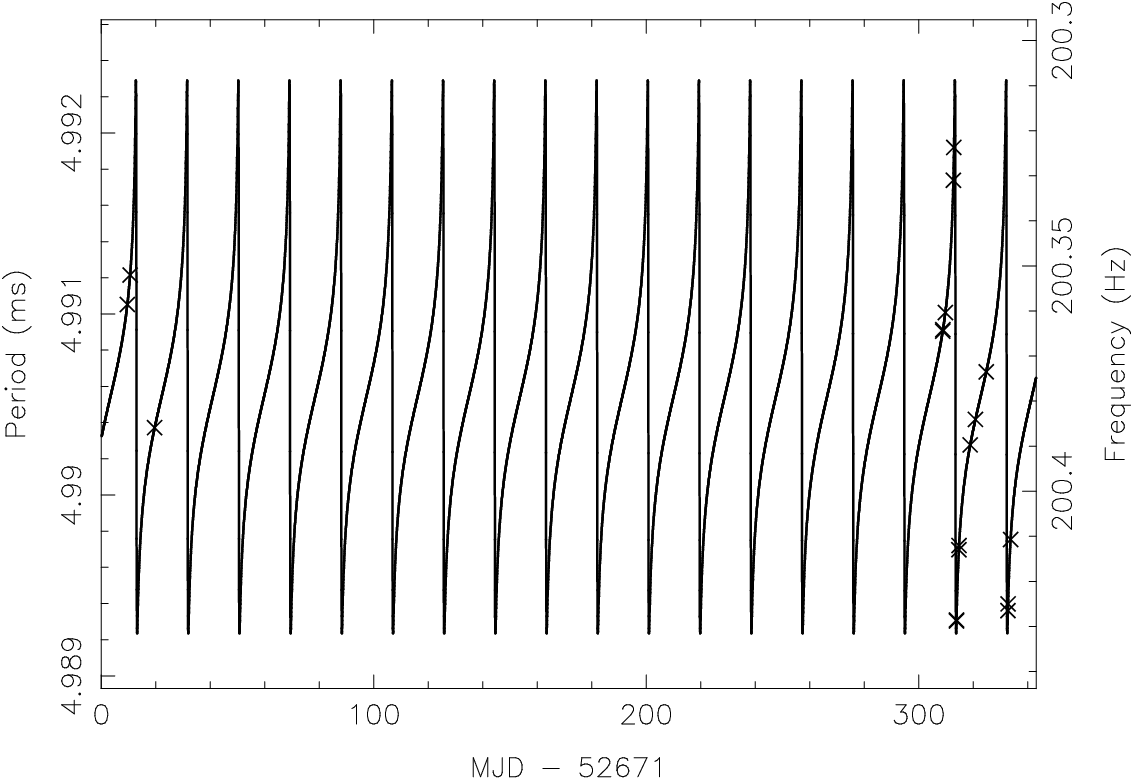}}
\end{picture}
\caption{The measured barycentric rotational periods of NGC~1851A as a function of MJD.}
\end{figure}

In the first scan for NGC~1851, taken on the 10 of February 2003, we
detected a clear pulsed signal with a period of 4.991 ms (see Figure
1) and a DM of 52.15(10)~cm$^{-3}\,$pc. Analysis of the rotational
periods using {\sc tempo}\footnote{http://pulsar.princeton.edu/tempo}
proved most surprising: the best-fit model (see Figure 2) indicates
$e\,=\,0.889(2)$, the most eccentric orbit of any known binary pulsar
and many orders of magnitude more eccentric than the typical MSP orbit
(for this and following quantities, the number in parenthesis
indicates the $1-\sigma$ uncertainty, which we conservatively estimate
to be ten times the formal value computed by {\sc tempo}, as there is
still no phase-coherent timing solution for this pulsar). The
orbital period $P_b$ is 18.7850(8) days, and the
semi-major axis of the orbit projected along the line-of-sight ($x$) is
$36.4(2)$ light seconds. This implies a minimum companion mass of
0.9~M$_{\odot}$, assuming a pulsar mass of 1.35~M$_{\odot}$ (for the
median of expected inclinations, 60$^\circ$, the companion mass is
 1.1~M$_{\odot}$). The epoch of periastron $T_0$ is  MJD =
52984.46(2), the longitude of periastron $\omega$ is
$82(1)^\circ$ and the rotational period $P$ is 4.990576(5) ms.
Imaging with the GMRT, made during the search and
confirmation observations, has shown that the pulsar is located very
near the centre of the cluster (see Figure 3), just ouside the
0.06-arcminute core. The estimated flux density at 327~MHz is about
3.4$\pm$0.4 mJy.

\begin{figure}
\setlength{\unitlength}{1in}
\begin{picture}(0,5.0)
\put(-0.1,-1.0){\includegraphics{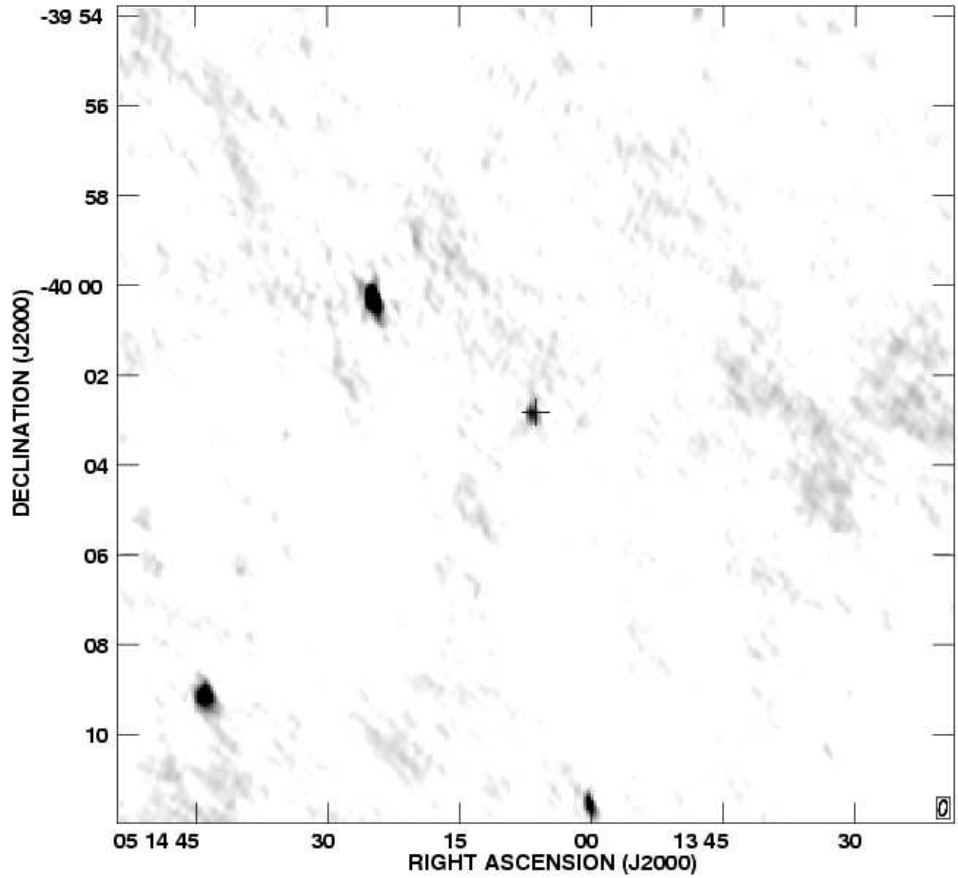}}
\end{picture}
\caption{A radio image of NGC~1851 at 327
MHz, made by combining the data from 4 different epochs of
observations in December 2003. The pulsar is the faint source near
the center of the cluster (indicated by the cross). The brighter
source to the north-east is not the pulsar, when the interferometer
was pointed at it no pulsed emission was detected. The small ellipse
in the lower right corner indicates the dimensions of the synthesized
interferometer beam, this interferometer includes {\em all} the
antennas in the array (the pulsation survey used only the central
square, which produces a 3-arcminute beam).}
\end{figure}

\section{Formation and nature}
\label{sec:formation}

 All known eccentric ($e > 0.1$) binary pulsars in the disk of the
Galaxy have relatively massive companions. This varied set of
companions includes blue giants, other neutron stars and heavy WDs.
Blue giants live only a few Myr which is probably not long enough to
allow the sustained mass accretion required to spin up a pulsar
companion to millisecond spin periods.  This is in accordance with the
observations, where the pulsars with massive companions have
rotational periods of tens or hundreds of milliseconds.
In addition, the second supernova event, where the giant becomes a
massive compact object, is likely to make the orbit significantly
eccentric, presuming the binary survives. Since both stars are now
compact, tidal circularization is henceforth impossible.  The
prolonged episode of stable mass accretion needed to spin up a neutron
star to millisecond periods is only possible from evolved lower-mass
MS stars. Such large timescales allow effective tidal orbit
circularization as well. This process likely created the MSP currently
in NGC~1851A, although with a presently unknown low-mass WD companion.
 
Some MSP $+$ low-mass white dwarf systems found in GCs, like
PSR~B1802$-$07 ($P = 23.1$\,ms, $P_b = 2.62$\,days, $e = 0.212$,
D'Amico et al. 1993) can become mildly eccentric due to interactions with
other objects in the cluster (Rasio and Heggie, 1995).
This is almost certainly not
the origin of the present NGC~1851A binary system since its
eccentricity is probably too high to be explained by this mechanism.
We are therefore lead to the conclusion
that after recycling, NGC~1851A exchanged its former low-mass
companion with a more massive object. This is the first system
presenting clear evidence of such a process. A massive star (the
pulsar's present companion) passed within a distance smaller than
about four times the separation of the components of the previous
binary system, a not uncommon event in the dense environment of a GC.
The most likely outcome from such an event is the formation of a
slightly tighter eccentric binary system containing the two more
massive objects (Hut 1996).

The companion of NGC~1851A has a minimum mass of 0.9\,M$_{\odot}$ and
its nature is as yet unclear, it could be either a compact or extended
object. In NGC~1851, a cluster with an age of $\sim$9\,Gyr (Salaris \&
Weiss 2002)
where 1-M$_{\odot}$ stars are now leaving the main sequence, such
objects should be readily detectable by the HST.  Because of the
lengthy episode of MSP recycling that preceded its formation, the
present NGC~1851A binary system is very likely to be a few Gyr younger
than the cluster in which it lies. Mathieu, Meibom and Dolan
(2004) have determined that, for the open cluster
NGC~188 (with an age of 7 Gyr and a stellar population similar to that
of GCs), binary systems containing MS stars with orbital periods
larger than 15 days have not yet had time to circularize. Therefore,
the observed eccentricity of the NGC~1851A system does not rule out
the possibility of the companion being an extended object.  In fact,
partial and/or irregular ``eclipses'' from an extended object such as
a MS star may explain the apparent flux variability from NGC~1851A.

\section{Conclusion}
 
We have discovered a remarkable 5-ms binary pulsar, the first to be
found in the GC NGC~1851 and the first pulsar to be discovered with
the GMRT. Its orbit is the most eccentric known for any system
containing a pulsar, while its rotational period is much shorter than
that of any other pulsar in an eccentric binary system.  This
combination of characteristics indicates that, after becoming an MSP
by accreting matter from a low-mass companion star, this neutron star
has almost certainly exchanged it for its present, significantly more
massive companion. This is the most effective way of forming a binary
system containing a millisecond pulsar and a black hole, provided
black holes exist in GCs. Follow-up studies of this object will allow
us to determine the nature of the companion and hopefully
measure the masses of both components of this binary system.

\end{document}